\documentclass[11pt]{article}
\usepackage{pst-all}
\usepackage{pstricks}
\usepackage{pstcol,pst-fill,pst-grad}
\usepackage{amsfonts}
\usepackage{graphicx}
\usepackage{amsmath}
\usepackage[normalem]{ulem}
\usepackage{multicol}
\usepackage{tikz}
\usetikzlibrary{matrix}
\usepackage{enumerate}
\RequirePackage{mathrsfs} \RequirePackage[sc]{mathpazo}
\RequirePackage{wasysym} \RequirePackage{setspace}

\makeatletter \@addtoreset{equation}{section}

\newcommand{\be}{\begin{equation}}
\newcommand{\ee}{\end{equation}}
\newcommand{\bea}{\begin{eqnarray}}
\newcommand{\eea}{\end{eqnarray}}
\begin{document}
\date{}
\title{   Multi-qubits and  Polyvalent  Singularity  in Type II Supestring Theory}
\author{ Adil Belhaj\thanks{belhaj@unizar.es}
\hspace*{-8pt} \\
\\
 {\small LIRST, D\'epartement de Physique, Facult\'e
Polydisciplinaire, Universit\'e Sultan Moulay Slimane}\\{ \small
B\'eni Mellal, Morocco } }\maketitle

\begin{abstract}
Inspired by geometric engineering method, we approach qubit systems
in the context of D-branes in type II superstrings. Concretely, we
establish   a correspondence between such  quantum systems and
polyvalent singularities appearing in local Calabi-Yau manifolds.
First, we examine 1-qubit by considering a D2-brane probing the
su(2) toric singularity associated with  type IIA  monovalent
geometry. Then, we discuss the multi-qubits in terms of $n$ factors
of su(2) singularities using the  Cartan decomposition of non zero
roots. Applying mirror symmetry, the 4-qubits  are linked to  the
tetravalent singularity associated with the affine $\widehat{so(8)}$
Lie algebra matching with  the ADE-correspondences in the context of
quantum information theory.  Precisely, these states  can be
identified with wrapped D4-branes in    a Calabi-Yau 4-fold near
such a   singularity.

{\bf Keys words }:  String theory;  Quantum information theory;
Graph theory;   Toric geometry; Calabi-Yau singularities; Lie
algebras.
\end{abstract}


\thispagestyle{empty}

\newpage \setcounter{page}{1}
\section{Introduction}
Quantum information theory (QIT)  has  attracted recently much
attention mainly in relation   with  black holes  and holography
\cite{1,2,3}. This theory, which has been considered as a possible
combination of computer science and quantum mechanics,   is based on
a fundamental piece called qubit. This object   has been dealt with
by applying  certain  mathematical operations associated with
tensor-product of  the Hilbert vector spaces\cite{4,5,6}.

Qubit systems  have been  extensively  studied  using different
approaches including  superstring models  and   graph theory [8-16].
 More precisely,
a nice  link  between the stringy  black holes and  qubit systems
have been  investigated by considering   the  compactification
scenario. More concretely, the  supersymmetric  STU black hole
derived from the II superstrings  has been linked  to 3-qubits using
the hyperdeterminant concept \cite{8,9}. This link, which  is known
by black hole/qubit correspondence, has been enriched  by many
extensions and  generalizations  to superqubits using supermanifold
calculations\cite{13}. In particular,   a study based on toric
complex geometry has been done  leading to   a classification of
qubit systems in terms of black holes in type II superstrings  using
D-branes \cite{16}.

 Alternative  investigations  have
conducted using graph theory including Andinka  which has been
explored  in the study of the supersymmetric representation theory.
These graphs have  been used to classify a class of
 qubits in terms of   extremal   black   branes in type II superstrings \cite{13}.
 Moreover,
 colored toric graphs of a product of  $\mathbb{CP}^1$ projective
 spaces  have been also used to deal with concepts of QIT including
 logic  gates \cite{15}.

The present work aims to contribute  to these activities  by
establishing a correspondence  between   multi-qubits   and
Calabi-Yau polyvalent  singularities  inspired by geometric
engineering method. Concretely,  these qubit systems  are
interpreted in terms D-branes of type II superstrings probing such
singularities. First, we examine 1-qubit by considering a D2-brane
probing su(2) toric singularity associated with type IIA monovalent
 geometry.  Then, the Cartan decomposition of non zero roots has allowed us to
discuss the multi-qubits  using $n$ factors of su(2)  toric
singularities. Applying mirror symmetry, the 4-qubits are linked
with the tetravalent singularity associated with the affine
$\widehat{so(8)}$ Lie algebra matching with  the ADE-correspondences
in the context of QIT.   Precisely, these states  can be identified
with wrapped D4-branes in    a Calabi-Yau 4-fold near such a
singularity.

 The organization of this
paper is as follows. In section 2, we give a concise  presentation
of polyvalent singularities explored in the geometric engineering
method.  Section 3 concerns lower dimensional qubits   in terms of
D2-branes  living  in type IIA superstring.  In section 4,  a
correspondence between multi-qubits and polyavent  singularities
appearing in local Calabi-Yau manifolds is established  by combining
D-branes, graph theory,  and Lie algebras. Section 5 is devoted to
discussions, open questions and speculations supported by Calabi-Yau
singularities.

\section{Polyvalent geometry in type IIA superstring}
A nice framework  to discuss polyvalent geometry  is  toric
description  of  complex manifolds  explored in type  II superstring
compactifications \cite{17}. It is recalled that a $n$-dimensional
toric manifold $\cal \bf M^n$ can
 be  represented  by a  toric
    polytope    $ \Delta({{\cal \bf V}^n})$ spanned by  $ k=n+r$
     vertices $ v_i$  belonging to the  $n$- dimensional lattice $\bf Z^n$  \cite{18,19,20}. These     vertices
     satisfy  $r$ relations
      \begin{equation}
    \sum \limits _{i=1}^{n+r} q_i^a v_ i=0,\quad a=1,\ldots,r,
  \end{equation}
where $q_{i}^{a}$ are integers called  Mori vectors carrying many
physical and mathematical data. The building model is the one
dimensional projective
 $\mathbb{CP}^1$ which    can be represented by two vertices
  $v_1$ and $v_2$ on the real line.  These two   vertices   verify   the following condition
\begin{equation}
\label{toriccp1}
v_1+v_2=0,
\end{equation}  which corresponds to
 the north and  the south poles of $\mathbb{CP}^1$,  respectively.  In this case, the   toric realization  is just
  the segment $[v_1,v_2]$ linking  the two  vertices  $v_1$ and $v_2$.
 For higher dimensional geometries, the toric  descriptions  are
slightly more complicated and can be found in \cite{18,19,20}.   It
has been understood that  toric geometry can be considered as  a
good place to study mirror symmetry discussed in the
compactification  of type II superstrings on Calabi-Yau manifolds
\cite{21,22,23}. In  such compactifications,  the  mirror symmetry
has been explored to study the  deformation of  the ADE
singularities based on  bivalent and trivalent geometries  appeared
in the toric realization of Calabi-Yau manifolds  used in the
     geometric engineering method of gauge theories \cite{17}.
 Following
\cite{21,22,23}, the mirror  geometry   can  be  obtained using such
a
 symmetry in the   associated sigma model  language.  In this
way, the  mirror geometry  is typically given  by a superpotential
in the dual Landau Ginzburg  (LG)   model \bea
 W&=&P(x_1,\ldots,x_{n-1})=\sum_{i}{y}_i,
\\ \nonumber
W&=&uv \eea subject to  \be
 \prod_{i=1}^{r+n} {y}_i^{q^a_i}=1,\ \ \ \ \ \ \ a=1\ldots,r .
\label{ta} \ee These equations have been handled to discuss the
  polyvalent  singularities   considered as an
extension  of the bivalent
    and trivalent geometries  used  in the geometric engineering method of   supersymmetric theories in terms of
    D-branes probing toric singularities \cite{17}. Before  working out  a general
picture of such a geometry,  it  is
    useful   to give a concise representation of lower dimensional
    cases
    involving   bivalent and trivalent ones.

The leading example is the monovalent geometry appearing in the
$A_2$ Dynkin diagram, where each vertex is connected only  to
another one. One of them is considered as the central vertex
represented by the following Mori  vector \be q^a_i=(-2,1,0,...,0).
\ee

 The
next  example  is the bivalent  model  known as linear geometry.
This bivalent vertex appears in the mirror geometry of type IIA
superstring on the $A_n$ ($n\geq 3$) space family. In the geometric
construction of supersymmetric quantum field theories, this  vertex
allows one  to build  a linear chain of gauge groups $ \prod SU $
with bi-fundamental matter fields. In particular, it has been used
to recover results of the Seiberg-Witten model from the  type IIA
superstring moving on  a K3 surface fibered  over on a chain of
$\mathbb{CP}^1$ curves  according to the   $A_n$ Dynkin
graphs\cite{17}. In toric geometry langauge, the corresponding Mori
vectors take the following form \be q^a_i=(-2,1,1,0,...,0). \ee In
local geometries of the K3 surface known by the deformed ALE spaces,
the bivalent vertices   represent  a  linear chain of divisors with
self intersection $(-2)$ and
 intersect  two adjacent divisors once with contribution $(+1)$.

 The trivalent geometry, however,  involves both bivalent and trivalent
vertices  which
 has appeared  in different occasions. In string theory for instance,  this geometry has been  used to  incorporate  fundamental
 matters in the geometric  construction of  a linear chain of $ \prod SU$ gauge groups
 \cite{17}.  Precisely, it helps to   examine  the  affine ADE singularities
 explored in the study of the elliptic   fibration of the  local  Calabi-Yau threefolds  which produce  $
N=2$ superconformal theories in four dimensions.
     In toric realization of the ALE spaces,  the trivalent geometry contains a central divisor with self intersection $(-2)$ intersecting
      three other divisors once with contribution $(+1)$. The
      corresponding Mori vector reads as
\be q_i=(-2, 1, 1, 1,0,...,0). \ee  The Calabi-Yau condition
requires that  this Mori vector should be modified  and takes the
following form \be q_i=(-2, 1, 1, 1,-1,0,...,0). \ee In type IIB
superstring theory, the mirror geometry  can be obtained by solving
the constraint given in the equation (2.4). Indeed, one finds the
following monomials \be
 1,x_1,x_2,x_3,x_1x_2x_3.\ee
 In this solution,  $1$ corresponds to central
divisor, while  $x_1$, $x_2$ and $x_3$  are associated with  the
remaining ones  having  contributions (+1). Here  the terms
$x_1x_2x_3$ corresponds to  an auxiliary divisor  with  contribution
$(-1)$ introduced to cancel the first class of Chern  required  by
the Calabi-Yau condition. A close inspection shows that the
trivalent geometry is relevant in the study of the complex
deformation of the $T_{p_1,p_2,p_3}$  trivalent singularity defined
as the intersection of three chains type $A_{p_1-1}$, $A_{p_2-1}$
and  $A_{p_3-1}$ appearing in the blowing up of elliptic exceptional
singularities $E_{6,7,8}$ \cite{17}. They are given by  the
following elliptic curves, respectively,
 \bea
 T_{3,3,3}: x_1^3+x_2^3+ x_3^3+  \lambda x_1x_2x_3\\ \nonumber
T_{2,4,4}: x_1^2+x_2^4+ x_3^4+  \lambda x_1x_2x_3\\
T_{2,3,6}:x_1^3+x_2^2+ x_3^6+ \lambda x_1x_2x_3 \nonumber \eea where
$ \lambda$ is a complex parameter.

In the fourthcoming sections, we establish a link between qubits and
polyvalent type IIA geometry using D-branes wrapping  on the
associated  spheres.

\section{ Lower dimensional qubits  and polyvalent geometries in type IIA superstring theory}
  To start,  it  is recalled that the physics of
qubit has been extensively investigated from
  different physical and  mathematical  aspects\cite{4,5,6,7}.   Using Dirac notation,  1-qubit is  described by the
  following state
\begin{equation}
|\psi\rangle=a_0|0\rangle+a_1 |1\rangle
\end{equation}
 where $a_i$  are complex  numbers   verifying the   probability
 condition
\begin{equation}
|a_0|^2+|a_1 |^2=1.
\end{equation}

It should be denoted that  this condition    can be interpreted
geometrically in terms of the so-called Bloch sphere,
$\mathbb{CP}^1$.  Similarly, the  2-qubits are represented by the
state
\begin{equation}
|\psi\rangle=a_{00}|00\rangle+a_{10}
|10\rangle+a_{01}|01\rangle+a_{11} |11\rangle
\end{equation}
with the probability condition
\begin{equation}
|a_{00}|^2+|a_{10}|^2+|a_{01}|^2+|a_{11}|^2=1.
\end{equation}
 This equation, however,  defines   a   3-dimensional complex projective space
$\mathbb{CP}^3$ generalizing the Bloch sphere. This  analysis can be
extended  to $n$-qubits associated with  $2^n$ configuration states.
Using the binary notation,  the  general state reads as
\begin{equation}
\label{qudit} |\psi\rangle=\sum\limits_{i_1\ldots i_n=0,1}a_{
i_1\ldots i_n}|i_1 \ldots i_n\rangle,
\end{equation}
where $a_{ i_1\ldots i_n}$  verify  the  real normalization
condition
\begin{equation}
\label{pcn} \sum\limits_{i_1\ldots i_n=0,1}a_{ i_1\ldots
i_n}\overline{a}_{ i_1\ldots i_n}=1
\end{equation}
defining   the  $\mathbb{CP}^{2^n-1}$ complex projective space. An
inspection shows that the qubit systems can be represented by
polyvalent  geometry using  type II D-branes. In what follows, we
refer to it as $n$-valent geometry. More precisely,  we will show
that  this is associated with the states describing the $n$-qubit
systems. In this way, a quantum state is interpreted in terms of
D2-branes wrapping spheres  in  type IIA superstring
compactifications. We expect that the $n$-valent geometry should
encode certain data  on the corresponding brane physics offering  a
new take on the graphic  representation of QIT using techniques
based on a combination of toric geometry and graph theory. To
understand how such a link could be true, we first examine  the case
of 1-qubit. Then, we give a general statement  for the $n$-valent
geometry.  For the 1-qubit, the  geometry is a local description of
the K3 surface where the manifold develops a su(3) singularity.  It
corresponds to vanishing two intersecting 2-spheres. Near such a
singular point, the K3 surface can be viewed as an  asymptotically
locally Euclidean (ALE) complex space which is algebraically given
by   the blowing down curves of the $A_2$ singularity. This
singularity can be removed by two intersecting $\mathbb{CP}^1$
curves according to the topology of the $A_2$ Dynkin graph.  This
picture  is considered as a 1-valent geometry since each
$\mathbb{CP}^1$ intersects only another one which can be represented
by two vertices in  the dual type IIB superstring theory. Deleting
such a vertex, we get one vertex corresponding to the $A_1$ Dynkin
graph associated with the toric $su(2)$ singularity of the  K3
surface. The local geometry of this background is described by the
algebraic  complex equation
$$ xy=z^2$$ where  $x,y,z$ are complex coordinates  of
 type IIA geometry. The  singularity  can  be deformed  by   blowing  up   the singular point
 by a  $\mathbb{CP}^1$  complex curve. In this  way,  a   D2-brane wrapping around such a
 complex curve  gives two   states $|\pm\rangle$  depending on  the two possible
      orientations for the  wrapping procedure. In Lie algebras,
      this  mapping can be supported
   by  the root system  decomposition of $ su(2)$
      symmetry given by
 \be     su(2)=h\oplus
g_{+\alpha}\oplus g_{-\alpha} \ee
 where $h$ is the associated Cartan subalgebra \cite{24,25,26}. Concretely,   these  two   states $|\pm\rangle$  correspond to   the
two dimensional vector space associated with the two roots
$\{\pm\alpha \}$
 \be
E_\alpha=\frac{su(2)}{h}=g_{+\alpha}\oplus g_{-\alpha}.
 \ee
Now, we consider the following corresponding
\begin{eqnarray}
g_{+\alpha}: |+\rangle  \leftrightarrow |0\rangle  \\
\nonumber g_{-\alpha}: |-\rangle  \leftrightarrow |1\rangle
\nonumber
 \end{eqnarray}
which gives a mapping between  qubit states   and the  states of the
D2-brane wrapping configuration space. In this case, the probability
of measuring the qubit in   certain state could be  determined  in
terms of winding numbers on $\mathbb{CP}^1$.  Moreover, it should be
noted that the Weyl group of $ su(2)$ Lie algebra which is
$Z_2=\{e,\sigma\}$ symmetry  can be associated with the Pauli $X$
gate,  acting  as a NOT gate. In two dimensional representation of
$Z_2$, one has
\begin{eqnarray}
\sigma=X=\left(%
\begin{array}{cc}
  0 & 1 \\
  1 & 0 \\
\end{array}%
\right).
 \end{eqnarray}

  Having  discussed
1-qubit case, we move now to the next model associated with the
2-valent geometry dealing with  the 2-qubit systems defined in a 4
dimensional Hilbert space.

The 2-qubit model, which is  interesting from  entanglement
applications,  will  be associated with the 2-valent geometry
appearing in the $A_3$ type IIA  superstring.  It  contains  a
central 2-sphere $\mathbb{CP}^1$ which intersects  two other ones
according to the $A_3$ Dynkin diagram in type IIB mirror
description. Removing the central vertex, we get a graph which can
be identified with the Dynkin
  graph of the $su(2)\oplus su(2)$ Lie algebra.

In this way,  the corresponding  type IIA geometry involves two
isolated $\mathbb{CP}^1$'s.  Two   D2-brane wrapping around such a
 geometry gives four   states  $|\pm\rangle \otimes  |\pm\rangle $  depending  on  the two possible
      orientations  on each   $\mathbb{CP}^1$. In Lie algebra formwork,
      these configurations correspond to   the four  non zero roots
      $\{\pm\alpha^i,\;i=1,2\}$  of  the    $su(2)\oplus su(2)$ Lie algebra.   This  can be  supported
by   the root system
     decomposition  providing a  four dimensional  vector space
 \be
E_\alpha=E_\alpha^1  \oplus  E_\alpha^2
 \ee
 where the factor  $E_\alpha^i$ is given by
  \be
E_\alpha^i=g_{+\alpha^i}\oplus g_{-\alpha^i}, \qquad i=1,2.
 \ee
  In this case, the Weyl group
will be identified with the   $Z_2\times Z_2$ symmetry associated
with the existence of   four  states in type IIA superstring using
D2-branes wrapping  on  two isolated $\mathbb{CP}^1$'s.

\section{Type IIA polyvalent  geometry of multi-qubits}
An inspection shows that  the  $n$-valent geometry can have also  a
nice graph theory realization  by combining  toric geometry and Lie
algebra structures.  Indeed, a graph  $G$ is  represented by a pair
of sets $G = (V(G), E(G))$, where $V(G)$ is  the vertex set and
$E(G)$ is associated with  the edge set \cite{27,28,29}. It is
recalled that two vertices are adjacent if they are connected by a
link. For any graph $G$, we define   a symmetric squared  matrix
called an adjacency matrix $I(G) = (I_{ij})$, whose elements are
either $0$ or $1$
\begin{equation}
I_{ij}=\left\{%
\begin{array}{ll}
    1, & \hbox{  $(i,j)\in E(G)$}, \\
    0, & \hbox{  $(i,j)\not\in E(G)$}. \\
\end{array}%
\right.
\end{equation}
This matrix, which  plays a primordial role  to
  provide connections with many areas in mathematical and physics,
 encodes  all the information residing on the graph. These data can be used to give a geometric representation  for  complicated
  systems  including   standard like models  of particle physics,
  or more generally  quiver gauge models  built from  string theory.
A special  example of  graphs being relevant  in the  present work
is called star graph formed by  a central  vertex that is connected
to other outer vertices. In this way,   the $n$-valent  geometry can
be represented by  a  star graph  containing  a cental vertex
connected with $n$ ones. In Lie algebra theory, these graphs could
represent   indefinite Lie algebras generalizing the finite and
affine symmetries. It is  observed that   for  $n=3$, we recover the
trivalent geometry appearing in  the finite $so(8)$ Lie algebra,
known also by  $D_4$. For $n=4 $,  however, we get the tetravalent
vertex of  the affine  $\widehat{ so(8)}$  Dynkin graph, known also
by $\widehat{D_4}$.  More generally, the  graph   which  represents
the $n$-valent geometry could  be worked out to  extend  the
$T_{p_1,p_2,p_3}$ trivalent singularity to $T_{p_1,p_3,\ldots,p_n}$
polyvalent singularities by considering  a non trivial intersection
of $n$ chains of bivalent geometries  formed by  spheres in the
Calabi-Yau manifolds. These geometries go beyond  the bivalent and
trivalent ones explored to engineer $N=2$ field models in four
dimensions from type II superstring compactifications \cite{17}.

Roughly, we would like to extend the results presented in the
previous section to  the $n$-valent geometry.
 Concretely, we can do something similar for $n > 2$.  In type IIA superstring,  associated with  the middle-degree cohomology, a  $n$-valent
geometry  can be described by  a central
 sphere $C_0$,   with self intersection $(-2)$ intersecting $n$ other
spheres  ($ C_i,i=1,\ldots,n$) with contribution $(+1)$.  The
corresponding intersection numbers  read as \bea
C_0.C_0&=&-2\nonumber\\
C_0. C_\ell &=1& \qquad \ell=1,\ldots,n. \eea In relation with the
adjacency matrix  in  graph theory and toric geometry, the
intersection numbers  of the  $n$-valent geometry  can be written as
follows \bea C_i.C_j&=&q_i^j\\\nonumber &=&-2\delta_{ij}+I_{ij}.\eea

However, the  Calabi-Yau condition requires that one should  add an
extra  non  compact cycle  $C_{n+1}$   with contribution $2-n$. In
string theory compactification on local Calabi-Yau  manifolds, this
cycle does not affect the corresponding physics.  In this way, the
Mori vector  representing the  central  sphere  should read as \be
q^0_i=(-2,\underbrace{1,\ldots, 1}_n, 2-n). \ee As in graph
 theory approach,  if we  delete  the    central  vertex, we get  the
 so-called empty graph having $n$ vertices. An inspection shows that
 this graph corresponds to  the Dynkin diagram   of a
 particular Lie algebra defined by $n$  pieces  of $
 su(2)$  denoted by  $\underbrace{su(2)\oplus \ldots \oplus
 su(2)}_n$. In this way,   there are  $2^n$ ways  for $n$    type II D-branes  to  wrap $n$ distinguishable blowing up spheres   in the type II
 superstrings.  It is remarked  that the  $su(2)$  singularities are distinguishable   giving  rise $2^n$
possible inequivalent  states $|\pm\rangle \otimes \ldots  \otimes
|\pm\rangle $. The associated root system  decomposition is
 \be
E_\alpha=E_\alpha^1  \oplus  \ldots  \oplus E_\alpha^n
 \ee
 where  $
E_\alpha^i=g_{+\alpha^i}\oplus g_{-\alpha^i}$.  Besides the  root
decomposition, the proposed  link   can be supported by the
corresponding homotopy group given by \be \Pi_2(
\underbrace{A_1\oplus \ldots \oplus A_1}_n)=\underbrace{Z_2\times
\ldots \times
 Z_2}_n.\ee
 This  link  may offer a novel way to study  quantum
geometry  associated with quantum mechanical theory of strings and
D-branes wrapping non trivial  cycles in the Calabi-Yau manifolds.

\section{Discussions and open questions}
In this work, we have approached the multi-qubits in the context of
type II superstrings. In particular, we   have  interpreted the
multi-qubits in terms of polyvalent geometry appearing in the
Calabi-Yau manifolds used in the geometric engineering  method.
First, we have examined the 1-qubit in terms of monovalent geometry
in type IIA superstring.  The geometry is no thing but the $A_2$ ALE
space. Deleting one node associated with the  central sphere,  we
have obtained just one 2-sphere in which  a  D2-brane can wrap to
give two states depending on the two possible orientations   of the
2-sphere. This scenario reproduces  the states  of   an 1-qubit
system. This  has been supported by the Cartan decomposition of
su(2) Lie algebra producing  a two dimensional vector space
associated with the non zero roots.  Then, we have given a general
picture describing $n$-qubits in terms of polyvalent geometry using
methods based on star graph operations and the Cartan decomposition
of $\underbrace{su(2)\oplus \ldots \oplus
 su(2)}_n$   Lie algebra. In this way, the  $n$-qubits  correspond
 to $2^n$ non zero roots of such a Lie   algebra.

It  is worth noting that the $n=4$  could  match  with  the work
 dealing with the ADE correspondence in the context of QIF \cite{30,301,300}. In that work the 4-qubits are linked to the
 $D_4$ singularity. In our case, however,    these   systems
 are  associated with the   $\widehat{so(8)}$ affine Lie algebra. It has been
remarked that this singularity  can  appear in the mirror  of a
toric Calabi-Yau manifold where the  Mori vectors, up some details,
are given in terms of  the $\widehat{so(8)}$ Cartan matrix.
 Solving the toric data associated with   the Dynkin graph   of
$\widehat{so(8)}$ Lie algebra, we find the follwing the mirror
geometry \be P(x_1,x_2, x_3, x_4, v)= x_1^4+x_2^4+ x_3^4+ x_4^4+
x_1x_2x_3x_4+ w(x_1^2+x_2^2+ x_3^2+ x_4^2)+w^2.
 \ee
describing a quasihomogenous hypersurface in the weighted projective
space $\mathbb{WCP}^4_{1,1,1,1,2}$. This  manifold    can be
associated with the tetravalent singularity
 \be
 T_{4,4,4,4}=x_1^4+x_2^4+ x_3^4+ x_4^4+ \lambda x_1x_2x_3x_4
 \ee
corresponding to  a quartic in $\mathbb{CP}^3$ considered as a
particular  fibre  identified with a K3 surface. This could  produce
a  singular  Calabi-Yau  4-fold with a  K3-fibration developing a
tetravalent singularity.  Type  IIA  superstring on such  a
four-fold
 singularity  can provide a similar statement. In this
case, the compact part of the  local  deformed geometry  is,
topologically, homotopic to a bouquet  of four-spheres according to
the  $\widehat{so(8)}$  Dynkin graph. The 4-qubit  states can be
identified with wrapped D4-branes in  such  a  Calabi-Yau 4-fold
near the tetravalent singularity.

 In the general case, we expect
that   the $n$-valent geometry  ($n > 2$)  could appear  in the
$n-2$ dimensional Calabi-Yau manifold fibrations of $(2n-4)$-folds.
The fiber  Calabi-Yau manifolds can be considered as homogeneous
hypersurfaces in $n-1$-dimensional projective spaces.  The $n$-qubit
states  can be determined in terms  of    wrapped $(2n-4)$-branes in
a Calabi-Yau  $(2n-4)$-folds near a polyvalent singularity.

This work comes up  with many open directions and speculations. The
intersecting problem is the discussion of other quantum concepts
including  the entanglement using polyvalent  singularities of type
II superstring compactifications. We believe that the central
vertex may play a primordial role in the study of such a concept.
Using graph theory, it  should be interesting to compute the
entanglement entropies of some edges with respect to the rest of
 the star  graph which be  useful to better understand
the physical characteristics and mathematical structures of
entangled states from Calabi-Yau  polyvalent singularities.  It has
been observed that there  could be  possible directions for future
investigation in  connection with quadrality for supersymmetric
models based on mirror symmetry and toric geometry realization of
local Calabi-Yau manifolds  discussed more recently in \cite{31}.
This will be addressed elsewhere.\\
\\
\\ \textbf{Acknowledgements}: The author would like to thank the
Instituto de Fisica teorica (IFT UAM-SCIC)  in Madrid for its
support via the Centro de Excelencia Severo Ochoa Program under
Grant SEV-2012-0249.  He   is very grateful to Luis Iba\~{n}ez for
scientific supports. He thanks also the  departamento de  Fisica
teorica, Universidad de  Zaragoza,  for scientific helps,   and
Antonio Segui for discussions on related topics. He also
acknowledges
 the warm hospitality  of  Monta\~{n}ez and Naz families  durante his
travel in Spain and he thanks also Hajja Fatima (his mother) for
patience and supports.

\end{document}